\documentstyle[amssymb,amsmath,12pt]{article}
\begin{document}

\title{ {\bf Integration of the mKdV hierarchy with integral type of source  
}  }
\author{ {\bf  Shuo Ye \hspace{0.8cm} Yunbo Zeng}
\\ {\small\em Department of Mathematical Science, Tsinghua University, 
Beijing 100084, China}}
\date{}
\maketitle
\renewcommand{\theequation}{\arabic{section}.\arabic{equation}}

{\bf{Abstract}}

\hspace{0.4cm}
We investigate the mKdV hierarchy with integral type of source (mKdVHWS), 
which consist
of the reduced AKNS eigenvalue problem with $r=q$ and the mKdV hierarchy
with extra term of the integration of square eigenfunction.
First we propose a method to find the explicit evolution equation
 for eigenfunction of the auxiliary linear problems of the mKdVHWS.
Then we  determine the evolution
equations of scattering data  corresponding to  the mKdVHWS
which allow us to solve the equation in the mKdVHWS by inverse scattering
transformation.

\hskip\parindent


\hskip\parindent

\section{Introduction}
\parindent=0.8cm
\hspace{0.8cm}
The soliton equations with integral type of source have important physical
application, for example,  
the nonlinear Schr\"{o}dinger equation with
integral type of source (NLSEWS) is relevant to some problems of
plasma physics and solid state physics \cite{1}.
It was shown in \cite{2} that
the NLSEWS can be integrated by the inverse scattering method for
the Dirac operator. The key point of the application of the
inverse scattering method to integration of  the NLSEWS in \cite{2} is the 
use
of the determining relations playing the same role as different
operator representations of the Lax type of nonlinear evolution
equations integrable by various modifications of this method. Just
using the determining relations Mel'nikov obtained the evolution
equations for all the scattering data of the Dirac operator corresponding to
NLSEWS.
Similar method was used to investigate the KdV equation with
integral type of source (KdVWS) in \cite{3}. The reason for use of the
determining relations in [2,3] is that the evolution equation of
eigenfunction for eigenvalue problem corresponding to the NLSEWS and KdVWS
was not found. In
fact, establish of these determining relations and derivation of
the evolution equations for all scattering data in [2,3] are quite
complicated and required some skill. \par
In present paper we investigate
the new mKdV hierarchy with integral type of sources (mKdVHWS), which consist
of the reduced AKNS eigenvalue problem with $r=q$ and the mKdV hierarchy
with extra term of the integration of square eigenfunction.
We first present a method to construct the zero-curvature
representation for mKdVHWS by finding the explicit evolution
equation for eigenfunction of the auxiliary linear problrm for mKdVHWS.
 Then we present a way to determine the
evolution equation for the scattering data corresponding to the mKdVHWS
, which implies that the mKdVHWS can be integrated by
the inverse scattering method. Comparing with the method by using
determining relation in [2,3],  the method proposed in this
paper for determining the evolution equation of the scattering data
is quite natural and simple. This general method can be applied to
other (1+1)-dimensional soliton equations
with integral type of source.
\section{ The mKdV hierarchy with integral type of source}
\setcounter{equation}{0}

Consider the reduced AKNS eigenvalue problem for $r=q$ \cite{20}

\begin{eqnarray}
\left(\begin{array}{l}
\phi_1\\
\phi_2
\end{array}  \right)_x
=U \left(
\begin{array}{l}
\phi_1\\
\phi_2
\end{array}  \right),\hspace{0.4cm}U=\left(\begin{array}{lr}
-\lambda &q\\
q & \lambda
\end{array}  \right).
\label{3.1}
\end{eqnarray}
The adjoint representation of (\ref{3.1}) reads  \cite{21}\\
\begin{equation}
V_x =[U,V]=UV-VU. \label{1.2}
\end{equation}
Set
\begin{equation}
V=\sum _{i=0}^{\infty}\left(\begin{array}{lr}
a_i &b_i\\
c_i &-a_i
\end{array}  \right)\lambda^{-i}
\label{1.2a}
\end{equation} Eq. (\ref{1.2}) yields
$$a_{0}=-1,b_{0}=c_{0}=a_{1}=0,b_{1}=c_{1}=q$$
$$a_{2}=\frac{1}{2}q^{2},b_{2}=-c_{2}=-\frac{1}{2}q_{x},\cdots,$$
and in general
$$b_{2m+1}=c_{2m+1}=Lb_{2m-1},\hspace{0.4cm}b_{2m}=-c_{2m}=-\frac{1}{2}Db_{2m
-1},$$
\begin{equation}
a_{2m+1}=0,\hspace{0.4cm}a_{2m}=2D^{-1}qb_{2m}. \label{L}
\end{equation}
where $$L=\frac{1}{4}D^{2}-qD^{-1}qD,\hspace{0.4cm}
D=\frac{\partial}{\partial x},\hspace{0.4cm} DD^{-1}=D^{-1}D=1.$$
Set
\begin{equation}
V^{(2n+1)}=\sum _{i=0}^{2n+1}\left(\begin{array}{lr}
a_i &b_i\\
c_i &-a_i
\end{array}  \right)\lambda^{2n+1-i}
\label{vn}
\end{equation}
and take
\begin{eqnarray}
\label{tn} \left(\begin{array}{l}
\phi_1\\
\phi_2
\end{array}  \right)_{t_{2n+1}}
=V^{(2n+1)}\left(
\begin{array}{l}
\phi_1\\
\phi_2
\end{array}  \right).
\end{eqnarray}
Then the compatibility conditions of Eqs. (\ref{3.1}) and
(\ref{tn}) give rise to the mKdV hierarchy \cite{20}
\begin{equation}
q_{t_{2n+1}}=-2b_{2n+2}=Db_{2n+1}=D\frac{\delta H_{2n+1}}{\delta
q},\hspace{0.4cm}n=0,1,\cdots,
\end{equation}
where
$$H_{2n+1}=\frac{2a_{2n+2}}{2n+1}.$$
 Using (\ref{3.1}), we have
\begin{equation} \label{L1}
\frac{\delta\lambda}{\delta
q}=\phi_{1}^{2}-\phi_{2}^{2},\hspace{1.0cm}
L(\phi_{1}^{2}-\phi_{2}^{2})=\lambda^{2}(\phi_{1}^{2}-\phi_{2}^{2}).
\end{equation}
The eigenvalue problem (2.1) with $q$ vanishing rapidly as $|x|$
tends to infinity has not discrete eigenvalue. As proposed in [2,3,6,7], 
the mKdV hierarchy with integral type of
source is defined by
\begin{subequations}
\label{3.5}
\begin{equation}
q_{t_{2n+1}}=D[b_{2n+1}+\int _{-\infty}^{\infty}C(t,\zeta)(\phi
^{2}_{1}(x,t,\zeta )-\phi ^{2}_{2}(x,t,\zeta ))d\zeta ]
\label{3.5a}
\end{equation}
\begin{equation}
\label{3.5b}\phi_{1,x}=-i\zeta \phi _{1}+q\phi
_{2},\hspace{0.3cm}\phi_{2,x}=q\phi _{1}+i\zeta\phi
_{2}\hspace{1.0cm}\zeta\in(-\infty,\infty)
\end{equation}
\end{subequations}
 we assume $q(x,t_{2n+1})$ tends rather
quickly to zero as $x\rightarrow\pm\infty$. According to this
condition we assume that
\begin{equation}
\phi _{1}(x,t,\zeta)\sim a(t,\zeta)exp(-i\zeta x),\hspace{0.6cm}\phi
_{2}(x,t,\zeta)\sim b(t,\zeta)exp(i\zeta x),\hspace{0.4cm}x\rightarrow
-\infty \label{3.6}
\end{equation}
where $C=C(t,\zeta)$, $a$=$a(t,\zeta)$ and $b$=$b(t,\zeta)$ are
complex functions of $t\geq 0$ and $\zeta\in (-\infty,\infty)$.
Moreover we assume that the functions $C$, $a$ and $b$ are chosen
so that the right-hand side of equation (\ref{3.5}) determines the
function absolutely integrable over $x$ along the whole real axis.
One can easily verify that the requirement will certainly be
satisfied if the function $E$ and $\Gamma$ of the form as argued
in \cite{2}
$$
E=|C(t,\zeta)|[|a(t,\zeta)|+|b(t,\zeta)|]^{2}
$$
$$
\Gamma=|\frac{\partial}{\partial\zeta}[C(t,\zeta)a^{2}(t,\zeta)]|
+|\frac{\partial}{\partial\zeta}[C(t,\zeta)b^{2}(t,\zeta)]|$$ at
any $t\geq 0$ satisfy the condition
$$
\int_{-\infty}^{\infty}[E(t,\zeta)+\Gamma(t,\zeta)+\Gamma^{2}(t,\zeta)]d\zeta
<\infty$$

\section{ The Lax representation }
\setcounter{equation}{0} Following the method proposed in [6,7,8],
in order to find the zero-curvature representation  for
(\ref{3.5}), we first consider
\begin{subequations} \label{2.99}\begin{equation}
\label{2.99a}D[b_{2n+1}+\int _{-\infty}^{\infty}C(t,\zeta)(\phi
^{2}_{1}(x,t,\zeta )-\phi ^{2}_{2}(x,t,\zeta ))d\zeta ]=0
\end{equation}
\begin{equation}
\label{2.99b}\phi_{1,x}=-i\zeta \phi _{1}+q\phi
_{2},\hspace{0.3cm}\phi_{2,x}=q\phi _{1}+i\zeta\phi
_{2}\hspace{1.0cm}\zeta\in(-\infty,\infty)
\end{equation}
\end{subequations}
We can  obtain the Lax
 representation for (\ref{2.99}) by  using the adjoint
 representation (\ref{1.2}). According to (\ref{L}), (\ref{L1})
 and (\ref{2.99}), we may define
$$\tilde{a}_i=a_i,\tilde{b}_i=b_i,\tilde{c}_i=c_i,\hspace{0.6cm}i=0,1,\cdots,
2n,$$
$$\tilde{b}_{2n+2m+1}=\tilde{c}_{2n+2m+1}=L\tilde{b}_{2n+2m-1}=
-\int _{-\infty}^{\infty}(i\zeta)^{2m+2}C(t,\zeta)
[\phi _{1}^{2}(x,t,\zeta )-\phi _{2}^{^2}(x,t,\zeta )]d\zeta $$
$$\tilde{b}_{2n+2m+2}=-\tilde{c}_{2n+2m+2}=-\frac 12D\tilde{b}_{2n+2m+1}=
-\int _{-\infty}^{\infty}(i\zeta)^{2m+1}C(t,\zeta)
[\phi _{1}^{2}(x,t,\zeta )+\phi _{2}^{^2}(x,t,\zeta )]d\zeta $$
$$\tilde{a}_{2n+2m+2}=2D^{-1}q\tilde{b}_{2n+2m+2}=
\int _{-\infty}^{\infty}(i\zeta)^{2m+1}C(t,\zeta)
\phi _{1}(x,t,\zeta )\phi _{2}(x,t,\zeta)d\zeta$$
$$\tilde{a}_{2n+2m+1}=0,\hspace{0.6cm}m=0,1,\cdots,$$
Then \begin{eqnarray*} N^{(2n+1)}=\left(\begin{array}{lr}
A^{(2n+1)}&B^{(2n+1)}\\
C^{(2n+1)}&D^{(2n+1)}\end{array}\right)\equiv
\lambda^{2n+1}\sum_{k=0}^{\infty}\left(\begin{array}{lr}
\tilde{a}_k&\tilde{b}_k\\
\tilde{c}_k&-\tilde{a}_k
\end{array}\right)\lambda^{-k}+\left(\begin{array}{lr}
\theta&0\\
0&\theta \end{array}\right)\end{eqnarray*} where $\theta$ is some
constant and
$$A^{(2n+1)}=\sum_{k=0}^{2n}
a_{k}\lambda^{2n+1-k}+\theta+\int
_{-\infty}^{\infty}\frac{2(i\zeta)(i\eta )C(t,\eta)\phi
_{1}(x,t,\eta )\phi _{2}(x,t,\eta )}{(i\zeta )^{2}-(i\eta
)^{2}}d\eta$$
$$B^{(2n+1)}=\sum_{k=0}^{2n}
b_{k}\lambda^{2n+1-k}+\int
_{-\infty}^{\infty}\frac{i\zeta(i\zeta-i\eta)C(t,\eta)\phi_{2}^{2}(x,t,\eta)
-i\zeta(i\zeta+i\eta)C(t,\eta)\phi_{1}^{2}(x,t,\eta)}{(i\zeta
)^{2}-(i\eta )^{2}}d\eta$$
$$C^{(2n+1)}=\sum_{k=0}^{2n}
c_{k}\lambda^{2n+1-k}+\int
_{-\infty}^{\infty}\frac{i\zeta(i\zeta+i\eta)C(t,\eta)\phi_{2}^{2}(x,t,\eta)
-i\zeta(i\zeta-i\eta)C(t,\eta)\phi_{1}^{2}(x,t,\eta)}{(i\zeta
)^{2}-(i\eta )^{2}}d\eta$$
$$D^{(2n+1)}=-\sum_{k=0}^{2n}
a_{k}\lambda^{2n+1-k}+\theta-\int
_{-\infty}^{\infty}\frac{2(i\zeta)(i\eta )C(t,\eta)\phi
_{1}(x,t,\eta )\phi _{2}(x,t,\eta )}{(i\zeta )^{2}-(i\eta
)^{2}}d\eta$$  also satisfies the adjoint representation
(\ref{1.2}), i.e.
\begin{equation} N_{x}^{(2n+1)}=[U,N^{(2n+1)}],
\end{equation}
which, in fact, gives rise to the Lax representation of
(\ref{2.99}). Since (\ref{2.99}) is the stationary equation of
(\ref{3.5}), it is easy to find that the zero-curvature
representation for the mKdV hierarchy with integral type of source
(\ref{3.5}) is given by
\begin{equation}
U_{t_{2n+1}}-N_{x}^{(2n+1)}+[U,N^{(2n+1)}]=0,
\end{equation}
with the auxiliary linear problems
\begin{subequations}
\label{2.13}
\begin{equation}
\label{2.13a} \left(\begin{array}{l}
\psi_1\\
\psi_2
\end{array}  \right)_x
=\left(\begin{array}{lr}
-\lambda &q\\
q & \lambda
\end{array}  \right)
\left(
\begin{array}{l}
\psi_1\\
\psi_2
\end{array}  \right)=\left(\begin{array}{lr}
-i\zeta &q\\
q & i\zeta
\end{array}  \right)
\left(
\begin{array}{l}
\psi_1\\
\psi_2
\end{array}  \right)         ,
\end{equation}
where $\lambda=i\zeta$ and
$$\psi _{1,t_{2n+1}}(x,t_{2n+1},\zeta)=(A^{(2n+1)}+\theta )\psi 
_{1}+B^{(2n+1)}\psi _{2}
\equiv\sum _{k=0}^{k=2n}(a_{k}\psi _{1}+b_{k}\psi _{2})\lambda 
^{2n+1-k}+\theta\psi _{1}$$
$$+\int _{-\infty}^{\infty}\frac{i\zeta C(t,\eta)}{(i\zeta )^{2}-(i\eta 
)^{2}}
[2(i\eta )\phi _{1}(x,t,\eta )\phi _{2}(x,t,\eta )\psi _{1}
+(i\zeta -i\eta )\phi ^{2}_{2}(x,t,\eta)\psi _{2} -(i\zeta +i\eta
)\phi ^{2}_{1}(x,t,\eta)\psi _{2}]d\eta,$$
$$\psi _{2,t_{2n+1}}(x,t_{2n+1},\zeta)=C^{(2n+1)}\psi 
_{1}+(-A^{(2n+1)}+\theta )\psi _{2}
\equiv\sum _{k=0}^{k=2n}(c_{k}\psi _{1}-a_{k}\psi _{2})\lambda 
^{2n+1-k}+\theta\psi _{2}$$

$$+\int _{-\infty}^{\infty}\frac{i\zeta C(t,\eta)}{(i\zeta
)^{2}-(i\eta )^{2}}[(i\zeta +i\eta )\phi ^{2}_{2}(x,t,\eta)\psi
_{1}-(i\zeta -i\eta )\phi ^{2}_{1}(x,t,\eta)\psi _{1}$$
\begin{equation}
-2(i\eta
)\phi _{1}(x,t,\eta )\phi _{2}(x,t,\eta )\psi _{2}]d\eta.
\label{2.13b}
\end{equation}
\end{subequations}
In this way we find the explicit evolution equations of
eigenfunction $\psi$. Indeed, this kind of evolution equation of
eigenfunction was not obtained in [2,3].

\section{ Evolution equation for the reflection coefficients }
\setcounter{equation}{0}
    Now we can derive equations describing the evolution in time
    $t$ of the S-matrix elements. This can be made as follows.
     We define the eigenfunctions  
$f^{-}(x,\zeta)=(f^{-}_{1}(x,\zeta),f^{-}_{2}(x,\zeta))^{T}$,
      $\bar f^{-}(x,\zeta)=(\bar f^{-}_{1}(x,\zeta),\bar 
f^{-}_{2}(x,\zeta))^{T}$,
       $f^{+}(x,\zeta)=(f^{+}_{1}(x,\zeta),f^{+}_{2}(x,\zeta))^{T}$
     and $\bar f^{+}(x,\zeta)=(\bar f^{+}_{1}(x,\zeta),$ $\bar 
f^{+}_{2}(x,\zeta))^{T}$
          (here and hereafter the $"T"$ means transposition) for the equation 
(\ref{2.13a}),
     and the following asymptotics are fulfilled at any $\zeta\in 
(-\infty,\infty)$
\begin{subequations}
\label{3.77}
\begin{equation}
\label{3.77a} f^{-}(x,\zeta)\sim\left(\begin{array}{l}
1\\
0
\end{array}\right)e^{-i\zeta x},  \hspace{0.4cm}\bar 
f^{-}(x,\zeta)\sim\left(\begin{array}{l}
0\\
-1
\end{array}\right)e^{i\zeta x},  as\hspace{0.2cm} x\rightarrow -\infty
\end{equation}
\begin{equation}
\label{3.77b}
 f^{+}(x,\zeta)\sim\left(\begin{array}{l}
0\\
1
\end{array}\right)e^{i\zeta x},  \hspace{0.4cm}\bar 
f^{+}(x,\zeta)\sim\left(\begin{array}{l}
1\\
0
\end{array}\right)e^{-i\zeta x},  as\hspace{0.2cm} x\rightarrow +\infty
\end{equation}
\end{subequations} As is known, the functions $f^{-}(x,\zeta)$ and
$f^{+}(x,\zeta)$ admit an analytical continuation in the parameter
$\zeta $ into the upper half-plane Im$\zeta> 0$, and the functions
$\bar f^{-}(x,\zeta)$ and $\bar f^{+}(x,\zeta)$ admit an
analytical continuation in the parameter $\zeta $ into the lower
half-plane Im$\zeta <0$. It is easily seen that at any real
$\zeta\in (-\infty,\infty)$ the pair of functions $f^{-}(x,\zeta
)$ and $\bar f^{-}(x,\zeta)$ forms a fundamental system of
solutions to (\ref{2.13a}). Hence, we may define
\begin{subequations}
\label{3.88}
\begin{equation}
f^{+}(x,\zeta)=S_{12}(\zeta)\bar f^{-}(x,\zeta
)+S_{22}(\zeta)f^{-}(x,\zeta) \label{3.88a}
\end{equation}
\begin{equation}
\bar f^{+}(x,\zeta)=S_{11}(\zeta)\bar f^{-}(x,\zeta
)+S_{21}(\zeta)f^{-}(x,\zeta) \label{3.88b}
\end{equation}
\end{subequations}
where the quantities $S_{11}=S_{11}(\zeta )$,
$S_{12}=S_{12}(\zeta)$, $S_{21}=S_{21}(\zeta)$ and
$S_{22}=S_{22}(\zeta)$ are independent of $x$. Taking account of
(\ref{3.77}) and (\ref{3.88}) we get that at any
$\zeta\in(-\infty,\infty)$ the equality
\begin{equation}
\label{equality2}
S_{11}(\zeta)S_{22}(\zeta)-S_{12}(\zeta)S_{21}(\zeta)=1.
\end{equation}
Under the assumption that $q(x,t)$ vanishes rapidly as
$|x|\rightarrow\infty$, we have $$a_{0}=-1,
\hspace{0.2cm}b_{0}=c_{0}=0,
\hspace{0.2cm}lim_{|x|\rightarrow\infty}a_{j}=lim_{|x|\rightarrow\infty}b_{j}
=lim_{|x|\rightarrow\infty}c_{j}=0,
\hspace{0.2cm}j=1,2,\cdots,2n.$$ We denote the parameter $\theta$
in (\ref{2.13b}) corresponding to $f^{+}(x,\zeta)$ by $\theta
^{+}$ and $\bar{f}^{+}(x,\zeta)$ by $\bar{\theta}^{+}$,
respectively. Substituting $f^{+}(x,\zeta )$, $\bar
f^{+}(x,\zeta)$ into (\ref{2.13b}) respectively, we have
$$\frac{\partial f^{+}_{1}(x,\zeta )}{\partial t_{2n+1}}
=\{\sum_{k=0}^{2n}a_{k}(i\zeta)^{2n+1-k}+\theta ^{+}
-\oint_{-\infty}^{\infty}(\frac{\zeta }{\eta
-\zeta}+\frac{\zeta}{\eta +\zeta})H(\eta )d\eta $$
$$-\pi(i\zeta)H(\zeta)+\pi(i\zeta)H(-\zeta)\}f^{+}_{1}(x,\zeta )$$
$$ +\{\sum _{k=0}^{2n}(b_{k}(i\zeta )^{2n+1-k}
+\oint_{-\infty}^{\infty}\frac{\zeta }{\eta -\zeta}H_{1}(\eta
)d\eta+\oint_{-\infty}^{\infty}\frac{\zeta}{\eta +\zeta}H_{2}(\eta
)d\eta$$\begin{subequations} \label{1.11}
\begin{equation}
+\pi(i\zeta)H_{1}(\zeta)-\pi(i\zeta)H_{2}(-\zeta)\}f^{+}_{2}(x,t,\zeta
), \label{1.11a}
\end{equation}

$$\frac{\partial f^{+}_{2}(x,\zeta )}{\partial t_{2n+1}}
 =\{\sum_{k=0}^{2n}c_{k}(i\zeta)^{2n+1-k}
  -\oint_{-\infty}^{\infty}\frac{\zeta }{\eta -\zeta}H_{2}(\eta )d\eta
  -\oint _{-\infty}^{\infty}\frac{\zeta}{\eta +\zeta}H_{1}(\eta )d\eta$$
 $$-\pi(i\zeta)H_{2}(\zeta)+\pi(i\zeta)H_{1}(-\zeta)\}f^{+}_{1}(x,\zeta )$$
$$ +\{\sum _{k=0}^{2n}(-a_{k}(i\zeta )^{2n+1-k}
-\oint_{-\infty}^{\infty}(\frac{i\zeta }{i\zeta -i\eta}
-\frac{i\zeta}{i\zeta +i\eta})H(\eta )d\eta +\theta ^{+}$$
\begin{equation}
+\pi(i\zeta)H(\zeta)-\pi(i\zeta)H(-\zeta)\}f^{+}_{2}(x,\zeta ),
\label{1.11b}
\end{equation}

$$\frac{\partial\bar f^{+}_{1}(x,\zeta )}{\partial t_{2n+1}}
=\{\sum_{k=0}^{2n}a_{k}(i\zeta)^{2n+1-k}+\bar\theta
^{+}-\oint_{-\infty}^{\infty}(\frac{\zeta }{\eta
-\zeta}+\frac{\zeta}{\eta +\zeta})H(\eta )d\eta $$
$$+\pi(i\zeta)H(\zeta)-\pi(i\zeta)H(-\zeta)\}\bar f^{+}_{1}(x,\zeta )$$
$$ +\{\sum _{k=0}^{2n}(b_{k}(i\zeta )^{2n+1-k}
+\oint_{-\infty}^{\infty}\frac{\zeta }{\eta -\zeta}H_{1}(\eta
)d\eta +\oint_{-\infty}^{\infty}\frac{\zeta}{\eta
+\zeta}H_{2}(\eta )d\eta$$
\begin{equation}
-\pi(i\zeta)H_{1}(\zeta)+\pi(i\zeta)H_{2}(-\zeta)\}\bar
f^{+}_{2}(x,\zeta ), \label{1.11c}
\end{equation}

$$\frac{\partial\bar f^{+}_{2}(x,\zeta )}{\partial t_{2n+1}}
=\{\sum_{k=0}^{2n}c_{k}(i\zeta)^{2n+1-k}
  -\oint_{-\infty}^{\infty}\frac{\zeta }{\eta -\zeta}H_{2}(\eta )d\eta
  -\oint _{-\infty}^{\infty}\frac{\zeta}{\eta +\zeta}H_{1}(\eta )d\eta$$
 $$+\pi(i\zeta)H_{2}(\zeta)-\pi(i\zeta)H_{1}(-\zeta)\}\bar f^{+}_{1}(x,\zeta 
)$$
$$ +\{\sum _{k=0}^{2n}(-a_{k}(i\zeta )^{2n+1-k}-
\oint_{-\infty}^{\infty}(\frac{i\zeta }{i\zeta -i\eta}
-\frac{i\zeta}{i\zeta +i\eta})H(\eta )d\eta +\bar\theta ^{+}$$
\begin{equation}
-\pi(i\zeta)H(\zeta)+\pi(i\zeta)H(-\zeta)\}\bar f^{+}_{2}(x,\zeta
), \label{1.11d}
\end{equation}
\end{subequations}
where the integral $\oint$ is taken as the principal value, 
the quantities $\theta^{+},\bar\theta^{+}$ will be determined in
the next section, and
$$H(\eta )=C(t,\eta)\phi _{1}(x,t,\eta)\phi _{2}(x,t,\eta),$$
\begin{equation}H_{1}(\eta )=C(t,\eta)\phi _{1}^{2}(x,t,\eta
),\hspace{0.4cm}H_{2}(\eta )=C(t,\eta)\phi _{2}^{2}(x,t,\eta).
\label{HH1H2}\end{equation} As $x\rightarrow -\infty$, we find
that the following asymptotics are valid:
$$\oint_{-\infty}^{\infty}\frac{\zeta }{\eta -\zeta}H_{1}(\eta )d\eta\sim\pi 
(i\zeta )
C(t,\zeta)a^{2}(\zeta ,t)e^{-2i\zeta x},$$
$$\oint_{-\infty}^{\infty}\frac{\zeta }{\eta +\zeta}H_{1}(\eta )d\eta\sim\pi 
(i\zeta )
C(t,-\zeta)a^{2}(-\zeta ,t)e^{2i\zeta x},$$
\begin{equation}
\oint_{-\infty}^{\infty}\frac{\zeta }{\eta -\zeta}H_{2}(\eta
)d\eta\sim-\pi (i\zeta )C(t,\zeta)b^{2}(\zeta ,t)e^{2i\zeta x},
\label{1.13}
\end{equation}
$$\oint_{-\infty}^{\infty}\frac{\zeta }{\eta +\zeta}H_{2}(\eta )d\eta\sim-\pi 
(i\zeta )
C(t,-\zeta)b^{2}(-\zeta ,t)e^{-2i\zeta x},$$ Substituting
(\ref{3.88}) into (\ref{1.11}) and using (\ref{1.13}), as
$x\rightarrow -\infty$, we have
$$\frac{\partial S_{22}(\zeta)}{\partial t_{2n+1}}=\{-(i\zeta )^{2n+1}+\theta 
^{+}
-\oint_{-\infty}^{\infty}(\frac{\zeta }{\eta -\zeta}
+\frac{\zeta}{\eta +\zeta})h(\eta )d\eta$$
$$-\pi(i\zeta)h(\zeta)+\pi(i\zeta)h(-\zeta)\}S_{22}(\zeta)$$
$$-\{2\pi(i\zeta)h_{1}(\zeta)-2\pi(i\zeta)h_{2}(-\zeta)\}S_{12}(\zeta),$$
$$\frac{\partial S_{12}(\zeta)}{\partial t_{2n+1}}=\{(i\zeta )^{2n+1}+\theta 
^{+}
+\oint_{-\infty}^{\infty}(\frac{\zeta }{\eta -\zeta}
+\frac{\zeta}{\eta +\zeta})h(\eta )d\eta$$
\begin{equation}
+\pi(i\zeta)h(\zeta)-\pi(i\zeta)h(-\zeta)\}S_{12}(\zeta),
\label{100}
\end{equation}
$$\frac{\partial S_{21}(\zeta)}{\partial t_{2n+1}}=\{-(i\zeta )^{2n+1}
+\bar\theta ^{+}-\oint_{-\infty}^{\infty}(\frac{\zeta }{\eta -\zeta}
+\frac{\zeta}{\eta +\zeta})h(\eta )d\eta$$
$$
+\pi(i\zeta)h(\zeta)-\pi(i\zeta)h(-\zeta)\}S_{21}(\zeta),
$$
$$\frac{\partial S_{11}(\zeta)}{\partial t_{2n+1}}=\{(i\zeta )^{2n+1}
+\bar\theta ^{+}+\oint_{-\infty}^{\infty}(\frac{\zeta }{\eta -\zeta}
+\frac{\zeta}{\eta +\zeta})h(\eta )d\eta$$
$$-\pi(i\zeta)h(\zeta)+\pi(i\zeta)h(-\zeta)\}S_{11}(\zeta)$$
$$-\{2\pi(i\zeta)h_{2}(\zeta)-2\pi(i\zeta)h_{1}(-\zeta)\}S_{21}(\zeta),$$
where
$$h(\eta )=C(t,\eta)a(\eta ,t)b(\eta ,t)$$
$$h_{1}(\eta )=C(t,\eta)a^{2}(\eta ,t),\hspace{0.4cm}h_{2}(\eta 
)=C(t,\eta)b^{2}(\eta ,t)$$
One can easily see that if $C=0$ or $a=b=0$ then the resultant
system (\ref{100}) coincides with those equations which appear in
the case of the mKdV hierarchy without a source. Also one can
verify that system (\ref{100}) is consistent with equality (\ref{equality2}).
Using
(\ref{100}), we find that the reflection coefficients
\begin{equation}
R_{1}(\zeta
,t)=\frac{S_{11}(\zeta)}{S_{21}(\zeta)},\hspace{0.4cm}R_{2}(\zeta
,t)=\frac{S_{22}(\zeta)}{S_{12}(\zeta)}
\end{equation}
satisfies the equation
\begin{subequations}
\label{3.1222}
$$\frac{\partial R_{1}(\zeta)}{\partial t_{2n+1}}=2\{(i\zeta )^{2n+1}
+\oint_{-\infty}^{\infty}(\frac{\zeta }{\eta -\zeta}
+\frac{\zeta}{\eta +\zeta})h(\eta )d\eta$$
$$-\pi(i\zeta)h(\zeta)+\pi(i\zeta)h(-\zeta)\}R_{1}(\zeta)$$
\begin{equation}
\label{3.1222a}-\{2\pi(i\zeta)h_{2}(\zeta)-2\pi(i\zeta)h_{1}(-\zeta)\},
\end{equation}
$$\frac{\partial R_{2}(\zeta)}{\partial t_{2n+1}}=2\{-(i\zeta )^{2n+1}
-\oint_{-\infty}^{\infty}(\frac{\zeta }{\eta -\zeta}
+\frac{\zeta}{\eta +\zeta})h(\eta )d\eta$$
$$-\pi(i\zeta)h(\zeta)+\pi(i\zeta)h(-\zeta)\}R_{2}(\zeta)$$
\begin{equation}
\label{3.1222b}-\{2\pi(i\zeta)h_{1}(\zeta)-2\pi(i\zeta)h_{2}(-\zeta)\}.
\end{equation}\end{subequations}
Then, it follows from (\ref{3.1222}) that the evolution of the
reflection coefficients $R_1, R_2$ are influenced by the integral
type of source which is integration of the square eigenfunctions
belonging to the continuous spectrum of the spectral problem
(\ref{3.1}). For the case $r=q$, there is no discrete eigenvalue
for the spectral problem (\ref{3.1}) if the potential $q=q(x,t)$
tends rather quickly to zero as $|x|\rightarrow\infty$. The
evolution equations for the reflection coefficients are presented
by (\ref{3.1222}) which implies that the mKdV hierarchy with
integral type of source can be solved by the inverse scattering
method.

\section{Conclusion}
\hspace{0.8cm}By means of the reduced AKNS eigenvalue problem with
$r=q$ which has no discrete eigenvalue, we construct the mKdV
hierarchy with integral type of source(mKdVHWS). We propose a
method to find the evolution equation of eigenfunction
corresponding to mKdVHWS and further to determine the evolution
equation for scattering data which enable us to solve mKdVHWS by
Inverse Scattering Transformation. Comparing with the method for
determining the evolution equation for scattering data in [2,3],
our approach is quite natural and simple.
\par It should be noted that  the
reduced AKNS spectral problem for $r=-q$ may have the discrete
 eigenvalue. In this case, the right-hand side of equation
 (\ref{3.5a}) need to be added by the sum of square eigenfunctions of 
(\ref{3.5b})
 corresponding to the discrete
 eigenvalue. We will show  in the further coming paper that the mKdV 
hierarchy with
 these two kinds of sources can also be integrated by Inverse Scattering 
Transformation.

\end{document}